\documentclass[10pt]{article}

\usepackage{amsfonts}

\def\form#1{(\ref{#1})}
\def\'{``}
\def\l{\lambda}

\def\s{\sigma}

\def\al{\alpha}

\def\we{\wedge}

\def\L{{\mathcal L}}                   %densita` lagrangiana
\def\U{{\cal U}}                       %Lagrang.duale
  
\def\H{{\mathcal H}}                   %densita` Hamiltoniana
\def\C{{\mathcal C}} 
 
\def\V{{\mathcal V}}

\def\F{{\mathcal F}}

\def\Lm{\mathop{\Longrightarrow}\limits}

\def\Re{I\kern-.36em R}                     %R dei reali 

\newcommand{\be}{\begin{equation}}
\newcommand{\ee}{\end{equation}}
\newcommand{\ba}{\begin{eqnarray}}
\newcommand{\ea}{\end{eqnarray}}
\newcommand{\baa}{\begin{eqnarray*}}
\newcommand{\eaa}{\end{eqnarray*}}
\newcommand{\bat}{\be\left\{\begin{array}{l}}
\newcommand{\eat}{\end{array}\right.\ee}

\def\QDE{\rule{2.5mm}{2.5mm}}
\def\CVD{$\phantom{'}$\hfill\QDE}

\newtheorem{Theorem}{Theorem}[section]

\newtheorem{Remark}[Theorem]{Remark} %

\makeindex

\begin{document}

\title{\bf  Charges and Energy   in 
Chern--Simons Theories and Lovelock Gravity}
\author{ G. Allemandi\footnote{allemandi@dm.unito.it}, M.Francaviglia\footnote{francaviglia@dm.unito.it}, M. Raiteri\footnote{raiteri@dm.unito.it} \\
 Dipartimento di Matematica,
{ Universit\`a degli Studi di Torino}, \\
 Via Carlo Alberto 10,
10123 Torino, Italy }
\date{}

\maketitle

\begin{abstract}
Starting from the $SO(2,2n)$ Chern--Simons  form in $(2n+1)$
dimensions we calculate the variation of conserved quantities in
 Lovelock gravity  and  Lovelock--Maxwell gravity through the covariant
formalism developed in
\cite{EM}.  Despite  the technical complexity of the Lovelock Lagrangian 
we obtain a remarkably simple expression for the variation of the
charges ensuing from  the diffeomorphism covariance of the theory. The
viability of the result is tested in specific applications and the formal
expression for the entropy of Lovelock black holes is recovered.\\


\end{abstract}

\section{Introduction}

 Lovelock gravity 
can be considered as the most general extension of 
General Relativity to higher dimensions: in fact, even if it contains
higher curvature interaction terms, it
 is nevertheless governed  by  field equations  which are just of 
 the second order  in the metric field  \cite{lanczos,lov}. 
Far from being just a mathematical curiosity,
 Lovelock gravity as well as higher  curvature theories in general, have
received in the recent past a renewed interest  motivated by the 
hope of learning something about the nature of quantum gravity. As a
matter of facts, higher curvature interaction terms arise in the study of
 back-reactions  of quantum field energy as well as 
in the  low--energy limit
of string theory (see  \cite{ct,boul,stringa,myers}). \\

In this paper we address the problem to calculate  the charges
associated  to the diffeomorphism symmetries  of Lovelock gravity  in any
odd dimensional spacetime. To this end  we make use of 
the recipe worked out in \cite{EM}  to algorithmically define  (the
variation of) conserved quantities  directly from the equations of motion.
One of the major advantages of the formalism developed in \cite{EM} is
that it enables  to establish  the transformation rules acting on 
conserved quantities in  the transition from different, but
nevertheless on--shell equivalent, Lagrangian theories. This property is
well suited to tackle with our problem. It was indeed shown in
\cite{Bana5d,CF} that  a special case  of
Lovelock gravity (with a fixed choice of the Lagrangian coefficients)   in
$D=2n+1$ dimensions can be derived  from a Chern-Simon theory of the
group
$SO(2, 2n)$ (or $SO(1, 2n+1)$). Combining together this remarkable result
with the formalism
of \cite{EM}
for conserved quantities, we shall show that the calculation
of  conserved quantities for Lovelock gravity becomes rather easier 
if we first compute them  in the Chern--Simons framework  and we take
then  into account the suitable transformation rules  on the field
variables as well as on the infinitesimal generators of symmetries. Even
though we end up in this way  with an expression which holds true only
for a very special  case of Lovelock gravity, the contribution to
conserved quantities coming from each  higher curvature interaction term 
can be nevertheless isolated. This property allows one  to derive a
general formula for (the variation of) conserved quantities for
\emph{any} Lovelock theory, i.e. with completely arbitrary choice of the
 coefficients in front of each  curvature interaction
term.\\

Let us now analyse the problem in detail.
 The $D=2n+1$ dimensional Chern-Simons 
Lagrangian for the anti-de Sitter gauge group  $SO(2, 2n)$  
is written in terms of a gauge connection
\be
A=A_\mu^{AB}dx^\mu J_{AB}
\ee
where $\mu= 0,\dots, 2n$;  $A,B=0, \dots ,2n+1$; and 
$J_{AB}$ are the Lie algebra generators:
\be
[J_{AB}, J_{CD}]={1\over 2}\left\{-J_{AC} \eta_{BD}+J_{AD}
\eta_{BC}+J_{BC}
\eta_{AD}-J_{BD} \eta_{AC}\right\}
\ee
with $\eta_{AB}=(-1,1,\dots,1,-1)$.

The Lagrangian turns out  to be \cite{CF}: \footnote{In the sequel we
shall make use of  the notation $f(j^ky)$ to denote  a quantity ``$\, f\,
$''  depending on  the variables $y$ together with their derivatives up to
the order
$k$.}

\be\label{CS5dLa}
\begin{array}{rl}
L_1(j^1 A)= \kappa \; \int_0^1 dt <A\wedge (t dA+ t^2 A^2)^n >
\end{array}
\ee  
where $<\,\;\,>$ denotes the invariant form:
\be
<J_{A_1B_1} \dots J_{A_{n+1}B_{n+1}}>=\epsilon_{A_1B_1
\dots A_{n+1}B_{n+1}}
\label{commJ}
\ee
According to \cite{Bana5d} the constant  $k$ in \form{CS5dLa} is fixed
as: 
\be
\kappa= {l\over (D-2)!\Omega_{D-2}}, \qquad 
l=const \label{costanteacc}
\ee
where $\Omega_{D-2}$ is the area of the $(D-2)$--dimensional unit sphere
and $l$ has the dimension of a length.
 The Euler--Lagrange field
equations ensuing from  
\form{CS5dLa} turn out to be:
\be
{\delta L_1\over \delta  A^{A_{n+1}B_{n+1}}}={ \kappa}   
\epsilon_{A_1B_1
\dots A_{n+1}B_{n+1}} F^{A_1 B_1}\wedge\dots\wedge 
F^{A_nB_n} =0\label{eq1}
\ee 
where $F^{AB}= dA^{AB}+ A^A{}_C\wedge A^{CB}$ is the field strength of
the gauge connection.
Furthermore, 
the set of generators $J_{AB}$ can be split as $J_{AB}=(J_{ab}, P_a)$,
$a=0,\dots ,D-1$; namely: into the generators $J_{ab}$ of Lorentz 
rotations in
$D$ dimensions and the generators $P_a:=
J_{a, 2n+1}$ of inner translations (see for example \cite{Ach}, \cite{CF},
\cite{Wi1}).  Accordingly, the $SO(2,2n)$ gauge connection can be
decomposed as follows:
\be
A=\omega^{ab} J_{ab}+{1 \over l}e^{a} P_{a}\label{AWE}
\ee 
  This assumption  implies that the field strength in  turn decomposes
as:
\be
F=(R^{ab}+ {1 \over l^2}e^{a} \we e^{b} )J_{ab}+{1 \over l} T^a\,P_{a}
\label{FWE}
\ee 
where 
\be
R^{ab}=d\omega^{ab}+ \omega^{a}{}_{c}\wedge  \omega^{cb}
\ee
is the field strength of the $SO(1,2n)$ gauge connection $\omega^{ab}$,
while 
\be
T^a=d e^a +\omega^a{}_b\wedge  e^b
\ee
is the torsion $2$--form.

Substituting expressions \form{AWE} and \form{FWE} into \form{CS5dLa}
and taking into account that the only terms which do not
vanish are those of the kind  $<J_{a_1b_1} \dots
J_{a_n b_n}P_a>=\epsilon_{a_1b_1
\dots a_{n}b_{n}a}$
we can rewrite   the Chern--Simons Lagrangian in terms of the new
  dynamical fields $(\omega, e)$. Apart from boundary terms we
obtain:
\be
L_2(e, j^1\omega)=\kappa\sum_{p=0}^n c_p \L_p(e,j^1\omega)\label{CS5dl}
\ee
where
\ba
&&c_p={1\over (D-2p)  } {n\choose p}{1\over   l^{D-2p}} \\
&&\L_p(e,j^1\omega)=
\epsilon_{a_1 \dots a_D}
 {R}^{a_1 a_2}\wedge\dots\wedge  {R}^{a_{2p-1}a_{2p}}\wedge e^{a_{2p+1}}
 \wedge\dots\wedge e^{a_D}\label{latter}
\ea
The Lagrangian \form{CS5dl} turns out to be the sum  of dimensionally
continued  Euler densities
 \cite{Bana5d,CF,TZ} and we shall refer to it as the 
\emph{Euler--Chern--Simons Lagrangian}.
 Notice
that in the  expressions \form{latter} the ``vielbein'' $e^{a}$ appears
without derivatives and the gauge  connection $\omega^{ab}$
 enters only through its  field strength $R$. This remarkable property is
due to the fact that  all the terms involving explicitly the torsion
and/or the gauge  connection $\omega$ are pushed, through integrations by
parts,   into  boundary terms which are discarded in the expression
\form{CS5dl}. Notice also that, until field equations are not yet solved, 
no a priori rule exists  relating the fields $e^{a}$ and $\omega^{ab}$:
they are completely independent.\footnote{\label{footnote1}
Geometrically speaking we
are assuming  that  the Lagrangian $L_2$ is based  on a configuration
bundle $Y$  which is a product bundle $Y=Y_1\times_M Y_2$ with fibered
coordinates 
$(x^\mu,e^a_\nu, \omega^{bc}_\rho)$,  where $(x^\mu,e^a_\nu)$ and
$(x^\mu, \omega^{bc}_\rho)$ are fibered coordinates on
$Y_1$ and $Y_2$, respectively. The configuration bundle $Y$ is thence a
\emph{gauge natural} bundle admitting  a principal $SO(1,2n)$ bundle $Q$
as structure bundle \cite{libroFF,kolar}. Accordingly $Y_1$ is the
spin--frame bundle
\cite{Koslor} while
$Y_2=J^1Q/SO(1,2n)$ is the connection bundle; see
\cite{kolar,Sarda}}

 Varying the Lagrangian  
\form{CS5dl} with respect  to the independent fields $(e,\omega)$ we
obtain, respectively, the field equations:
\be
\label{emoti}
\cases{
{\delta L_2\over \delta e^a}= {\kappa\over l} 
\epsilon_{a a_1 b_1\dots a_n b_n} 
\hat{R}^{a_1b_1}\wedge\dots\wedge  \hat{R}^{a_nb_n} =0\cr
\cr
{\delta L_2\over \delta \omega^{a_nb_n}}= {n \kappa\over l} 
 \epsilon_{a a_1 b_1\dots a_n b_n} 
 T^a\wedge \hat{R}^{a_1 b_1}\wedge\dots\wedge\hat{R}^{a_{n-1}b_{n-1}}
=0 }\label{eqq2}
\ee
where
\be
\hat{R}^{ab}=R^{ab}+ {1 \over l^2}e^{a} \we e^{b} \label{12}
\ee
Notice that field equations \form{emoti} are dynamically
equivalent to field equations
\form{eq1} once the substitution \form{AWE} is taken into account.
Namely, the difference between the two sets of equations is just a
matter of notation. 
 
If we now consider  the particular solution  $T^a=0$  of the second set
of field equations \form{emoti}, it turns out that the $SO(1,2n)$
connection $\omega^{ab}$ is the spin connection of the vielbein $e^a$:
\be
\omega^{a}{}_{b\mu}= e^a_\alpha (\Sigma^\al{}_{\beta
\mu}-\Sigma_\beta{}^\alpha{}_\mu+ \Sigma_{\mu\beta}{}^\alpha  
)e^\beta_b\,
,
\qquad \Sigma^\al{}_{\beta
\mu}=e^\alpha_c\,\partial_{[\beta }\,e^c_{\mu]}\label{13}
\ee
and 
\be
R^{ab}{}_{\mu\nu}=e^a_\rho e^b_\sigma
R^{\rho\sigma}{}_{\mu\nu}(j^2g)\label{Riem}
\ee
where $R^{\rho\sigma}{}_{\mu\nu}(j^2g)$ is the Riemann tensor of the
metric $g_{\mu\nu}=\eta_{ab}e^a_\mu e^b_\nu$.
Inserting \form{Riem} back into the Lagrangian \form{CS5dl} we end up
with a special Lovelock Lagrangian, namely:
\be
L_3(j^2 g):=\left.L_2(e,j^1\omega)\right\vert_{T^a=0}=\kappa
\sum_{p=0}^n \bar c_p\bar \L_p(j^2g)\label{LGB}
\ee
where:
\ba
&&\bar c_p=  {n\choose p}{(2n-2p)!\over   l^{D-2p}}\label{CP} \\
&&\bar\L_p(j^2g)={1\over 2^p}\;\sqrt{g}\;
\delta^{\mu_1\dots \mu_{2p}}_{\nu_1\dots \nu_{2p}}
 {R}^{\nu_1 \nu_2}_{\mu_1 \mu_2}\dots{R}^{\nu_{2p-1}
\nu_{2p}}_{\mu_{2p-1} \mu_{2p}}\label{L11}
\ea
and $\delta^{\mu_1\dots \mu_{2p}}_{\nu_1\dots \nu_{2p}}$ 
denotes the totally skew--symmetric  product  of Kronecker deltas 
normalized to have values $0,\pm 1$.\footnote{In writing \form{L11} we
made use of the fact that, according to our notation, 
\be
\epsilon^{\mu_1\dots \mu_{2p}\alpha_{2p+1}\dots
\alpha_D}\; \epsilon_{\nu_1\dots \nu_{2p}\alpha_{2p+1}\dots
\alpha_D}=(D-2p)!\;\delta^{\mu_1\dots \mu_{2p}}_{\nu_1\dots \nu_{2p}}
\ee
}
In particular: $\bar\L_0=\sqrt{g}$ gives rise to the term corresponding to
a cosmological constant $\Lambda$ proportional to $-1/l^2$ (or
proportional to $1/l^2$ if we had started with  the group $SO(1,2n+1)$);
$\bar\L_1=\sqrt{g}\; R$ is the Hilbert Lagrangian in dimension $2n+1$ and
$\bar\L_2=\sqrt{g}\;(R^2-4R^\alpha_\beta R^\beta_\alpha +R_{\alpha
\beta}^{\mu \nu} R^{\alpha
\beta}_{\mu \nu})$
is 
 the Gauss--Bonnet term. Notice that
the coefficients in  the Lovelock Lagrangian \form{LGB} are uniquely
fixed and in fact dictated  by the initial  Chern--Simons form
\form{CS5dLa}; see \cite{Bana5d,CF,BHS}. From now on we shall refer to the
Lagrangian
\form{LGB} as
\emph{Lovelock--Chern--Simons Lagrangian}.\\

Let us now summarize. 
We have introduced so far three different Lagrangians; see
expressions
\form{CS5dLa}, \form{CS5dl} and \form{LGB}. The transition among them can
be schematically drawn as follows:
\be\
L_1(j^1A)\,\,\stackrel{A=A(\omega, e)}{\Longrightarrow}\,\,
L_2(e,j^1 \omega)
\,\,\stackrel{T=0}{\Longrightarrow} \,\,L_3(j^2 g)
\label{transition}\ee 
Notice however that the equivalence between the first two Lagrangians, as
already explained, holds true off--shell. On the contrary, the 
equivalence between $L_2$ and $\,L_3$ is just an on--shell
equivalence since it  holds true only along   the space of solutions with 
$T=0$.

The goal of the present paper is to calculate the conserved quantities,
in particular the energy,  for  each one of the Lagrangians in
\form{transition}. We point out that this task is more subtle than it
could first look like. The reason is mainly due to two different
causes. 
\begin{enumerate}
\item Let us
suppose that we  are somehow able to calculate the charge
$Q(L_{1},
\Xi)$ associated to the Lagrangian $L_1(j^1 A)$ and relative to an
infinitesimal generator of symmetries $\Xi$ which is common to both
the Lagrangians $L_1$ and $L_2$. According to the prescription
\form{transition} we could be induced to identify the charge  $Q(L_{2},
\Xi)$, relative to the second Lagrangian  in \form{transition} and
associated with the same generator $\Xi$, with the expression 
$Q(L_2,
\Xi)=\left.Q(L_1,
\Xi)\right\vert_{A(\omega, e)}$, i.e. by merely inserting the splitting
\form{AWE} into  $Q(L_1,
\Xi)$. This rule could  fail in giving the right expression. Indeed we
stress that in the transition from the Lagrangian \form{CS5dLa} to the
Lagrangian \form{CS5dl} divergence terms have to be discarded. As far as
field equations are concerned, those divergence terms are clearly
irrelevant. On the contrary, they can become strongly relevant when
dealing with conserved quantities. For example,  conserved
quantities calculated via   Noether
theorem, Hamiltonian or Hamilton--Jacobi--like methods  are all sensitive
to divergence terms of the  Lagrangian (see, e. g.,
\cite{Booth,BY,Lagrange,HawHun,Wald,RT}).

\item 
Let us now suppose we have been somehow able to bypass the aforementioned
problem. There  remain still another open question. Indeed, 
conserved quantities of a field theory are mathematical
expressions which are attributed  the meaning of physical observables.
They are built out by keeping into account the symmetry properties of the
field theory we are dealing with. Hence,  we have to know the 
relationships interplaying among  the symmetries of the three Lagrangians
\form{transition} (or  their field equations) if we want to establish a
correspondence among their charges, too. 
Entering into the details of the matter,
we observe that the Chern--Simons theory  is a \emph{gauge natural}
theory;  see \cite{libroFF,Koslor,Remarks,Godina,kolar}. It is based on a
configuration bundle
$\C$ which is the bundle of connections $\C=J^1P/SO(2,2n)$, where $P$
denotes a principal bundle  with Lie group $SO(2,2n)$, called  the
\emph{structure bundle } of the theory;
see \cite{Sarda}. Fibered coordinates on $\C$  are $(x^\mu, A^{AB}_\nu)$.
The most general
infinitesimal generator of symmetries for the equations of motion 
\form{eq1} is a projectable vector field $\Xi_P$ on $P$ of the kind
\be
\Xi_P=\xi^\mu\partial_\mu + \Xi^{AB} \rho_{AB}
\ee
where $\rho_{AB}$ is a local basis of right invariant vector fields on $P$
(in a trivialization
$(x,g)$ of $P$ we have
$\rho=(g\,\partial/\partial g)$, $g\in SO(2,2n)$).
 The
vector field $\Xi_P$  induces a transformation of the dynamical fields
according to the rule:
\be
\pounds_{\Xi_P} A^{AB}_\mu=\xi^\rho d_\rho  A^{AB}_\mu  + d_\mu \xi^\rho
A^{AB}_\rho + \stackrel{A}{D}_\mu \Xi^{AB}\label{18}
\ee
and pure gauge transformations correspond to $\xi^\mu=0$. Therefore the
charge $Q(L_1,
\Xi_P)$ associated to this symmetry must depend, in general, on the
dynamical field
$A$, on the coefficients $\xi^\mu$ and $\Xi^{AB}$ together with their
derivatives up to a fixed (finite) order. 
If we now consider  the splitting 
\be
\Xi^{AB}=(\Xi^{ab}, \Xi^a)
\ee
of the vertical part of the vector field $\Xi_P$ into ``rotational'' part
and ``translational'' part, we obtain from \form{AWE} and \form{18}
\ba
\pounds_{\Xi_P}\omega^{ab}_\mu&=&\xi^\rho d_\rho  \omega^{ab}_\mu  + d_\mu
\xi^\rho\omega^{ab}_\rho+\stackrel{\omega}{D}_\mu\Xi^{ab}+{1\over
l}e^a_\mu
\Xi^b -{1\over l}e^b_\mu \Xi^a \nonumber
\\
\pounds_{\Xi_P} e^a_\mu&=&\xi^\rho d_\rho  e^a_\mu + d_\mu
\xi^\rho e^a_\rho+
l\stackrel{\omega}{D}_\mu\Xi^{a}
-\Xi^a{}_c e^c_\mu
\label{poundso}
\ea
and, if we do not  assume a priori  the condition $T^a=0$, field equations
\form{emoti} are  \emph{separately} invariant 
 for the transformation \form{poundso}
only  when we set
$\Xi^a=0$.\footnote{Basically, the Chern--Simons  symmetry is broken  when
we declare that  the dynamical fields $(e,\omega)$ are independent fields
or, in other words,  when we assume that the configuration bundle
$Y=Y_1\times_M Y_2$ has a product bundle structure. From
\form{poundso} it is clear that ``inner translations'' of the AdS group,
namely those generated by the components
$\Xi^a$, induce transformations on the configuration bundle which do not
respect the product bundle structure.}

It follows that the most  general
infinitesimal generator of symmetries for each one of the equations of
motion 
\form{emoti} is a projectable vector field  $\Xi_Q$ on
the relevant $SO(2n,1)$ principal bundle $Q$ of the theory (see 
footnote \ref{footnote1}) which, in  turn, is a subbundle of $P$. In
coordinates:
\be
\Xi_Q=\xi^\mu\partial_\mu + \Xi^{ab} \rho_{ab}\label{21}
\ee
where now $\rho_{ab}$ is a local basis of right invariant vector fields on
$Q$. The charge
$Q(L_2,
\Xi_Q)$ associated to the vector field \form{21} must depend, in general,
on the dynamical fields
$(e,\omega)$, on the coefficients $\xi^\mu$ and $\Xi^{ab}$ together with
their derivatives up to a fixed order and can be \emph{suitable} obtained
from
$Q(L_1,
\Xi_P)$ through the prescription \form{AWE} on the dynamical variables and
the prescription $\Xi^a=0$ on the generators of symmetries.

Let us finally draw our attention to   the Lovelock--Chern--Simons
Lagrangian \form{LGB}. The latter Lagrangian is a
\emph{natural} (also said \emph{covariant} or   \emph{diffeomorphism
invariant}; see \cite{Robutti,Wald,kolar}) Lagrangian  and the generators
of symmetries are (not necessarily Killing) vector fields
$\xi=\xi^\mu\partial_\mu$ on the base manifold (i.e. the $D$--dimensional
spacetime). It follows that the charges $Q(L_3,
\xi)$  must depend, in general, on
the dynamical field
$g$ and on the coefficients $\xi^\mu$  together with
their derivatives up to a fixed order. To obtain a correspondence
between 
$Q(L_2,
\Xi_Q)$ and $Q(L_3,
\xi)$ the prescriptions \form{13} and \form{Riem} alone are not
sufficient since we also need a prescription to fix the coefficients
$\Xi^{ab}=\Phi^{ab}(\xi^\mu, d_\nu\xi^\mu, \dots)$ in terms of the
coefficients
$\xi^\mu$ and their derivatives up to a fixed order, namely we need to
select a 
\emph{lift} of the vector field $\xi$ up to the structure bundle $Q$
of the Lagrangian $L_2$. Notice that the rule 
$\Xi^{ab}=0$ is only local and it is not preserved (in general) by the
automorphisms of the structure  bundle $Q$, meaning  that we have to
resort to some other
\emph{covariant} lift. If we are able to select a lift
$\Xi^{ab}=\Phi^{ab}(\xi^\mu, d_\nu\xi^\mu, \dots)$ which is mathematically
well--defined (i.e. a covariant lift) as well as  physically viable
(i.e. it allows to reproduce the expected physical values for conserved
quantities in specific applications) we have the following scheme:
\be
Q(L_1(j^1A), \Xi_P)\,\,\Lm^{A=A(\omega,
e)}_{\Xi^a=0}\,\, Q(L_2(e,j^1\omega),\Xi_Q)
\,\,\Lm^{T=0}_{\Xi^{ab}=\Phi^{ab}(\xi) }
\,\,Q(L_3( g),\xi)
\label{transition2}
\ee 
governing the transition among conserved quantities.\\
\end{enumerate}

In the sequel we shall follow a recipe  to calculate conserved
quantities which has been recently worked out in \cite{EM}. The formalism
developed in \cite{EM} is in fact  well suited to tackle the issue we
are dealing with, mainly because  it allows to overcome both the  problems
we have just outlined.  Basically, the construction of conserved
quantities is  defined as follows. Given a field theory
the dynamics of which is derived  from a Lagrangian
$L$ and given a vector field $\Xi$ which is an infinitesimal generator of
symmetries for the equations of motions ${\delta L\over \delta y}$, one can
construct 
\emph{the variational Lagrangian } $L'(L,\Xi)$ through the contraction of
field equations with the Lie derivatives of fields, i.e.
$L'(L,\Xi)=-{\delta L\over \delta y^i}\pounds_\Xi y^i$. Handling the
variational Lagrangian with the tools of Variational Calculus in jet
bundles it is possible to algorithmically define the \emph{
variation} $\delta Q (L', \Xi)$ of \emph{covariantly} conserved quantities
along a one--parameter family of solutions.

Since the variation of conserved quantities is obtained  directly  from
the equations of motion the drawback related to problem  1) above is
automatically  overcome: the procedure is not affected by whichever
divergence term we add to a given Lagrangian since field equations are
insensitive to it. Moreover, also the problem raised in problem 2) can be
easily bypassed. Indeed, by imposing  the equivalence between  the 
variational Lagrangian
$L'(L_2,
\Xi_Q)$ (built out starting from   the equations of motion \form{emoti} 
and the generator of symmetries \form{21}) and the  variational Lagrangian
$L'(L_3,
\xi)$ (where $\xi$ is any spacetime vector field) a
preferred lift $\Xi_Q=\Phi(\xi^\mu, d_\nu\xi^\mu, \dots)$ of
spacetime vector fields is automatically ruled out from the theory and it
allows to complete the diagram \form{transition2}. The selected lift turns
out to be the so--called \emph{generalised Kosmann lift}; see 
\cite{Koslor, Godina, vangoden}.\\

Once we have understood the rules which allow us to make the transition
from the (variation of the) conserved quantities $\delta Q(L'_1,\Xi_P)$ 
ensuing from the Chern--Simons Lagrangian
\form{CS5dLa} and the (variation of the) conserved quantities
$\delta Q(L'_3,\xi)$ associated to the Lagrangian
\form{LGB}, the latter quantities can be easily computed from the former
ones. Moreover, even if the quantities obtained in this way are
relative to the whole Lagrangian \form{LGB}, the contribution coming from
each Lagrangian $\bar\L_p$  can be easily recognized and, therefore, we
can define the conserved quantities for a generic Lovelock Lagrangian
with arbitrary coefficients in front of each term. Hence, despite  the
technical complexity of the Lovelock Lagrangian,  we obtain  a
remarkably simple expression for its  charges.
\vspace{.6cm}

The rest of the paper is organized as follows.
In section \ref{Conserved Quantities} we shall shortly summarize the
formalism
developed in \cite{EM} to obtain the variation of conserved quantities
directly from the equations of motion. We then  apply it to calculate
conserved quantities relative to the Chern--Simons form
\form{CS5dLa}. Following the recipe schematically drawn in
\form{transition2} we  are then able to define the variation of conserved
quantities for Lovelock gravity. A straightforward generalization of the
aforementioned formalism will allow the generalization to 
Lovelock--Maxwell gravity.
In   section \ref{The spherically symmetric charged solution}
 the viability of the results previously obtained is tested
for a five--dimensional charged  black hole solution and the first law of
black hole mechanics directly follows from the homological properties
conserved quantities must satisfy.
 In
section \ref{entropia} 
 the formal
expression for the entropy of Lovelock stationary black holes  is finally
calculated  and it agrees with previous definitions; 
see \cite{Wald,T.Jacobson,visser}.
%%%%%%%%%%%%%%%%%%%%%%%%%%%%%%%%%%%%%%%%%%%%%%%%
\section{Conserved Quantities}
\label{Conserved Quantities}
In order to make the paper self--contained we shall here summarize the
formalism
to obtain the variation of conserved quantities
directly from the equations of motion.
Since  we are   interested only  in the application of the technique 
to the Chern--Simons theory and to Lovelock gravity,  we
shall skip the rigorous geometric framework and  we refer the reader to
\cite{EM} for a deeper insight into the
mathematical details of the matter. To stimulate  the interest of
physically--oriented  readers we shall keep as much as possible the
formal setting and the technical details to a minimum.

Let us consider  a  field theory described through a Lagrangian of order
$k$.
It can be locally  written as
$L=\L (y^i, \dots,y^i_{\mu_1\dots\mu_k})\, ds 
$
where  $\L$ is  the
Lagrangian density, 
$ds=dx^1\wedge\dots\wedge dx^D$ is the local volume element  of
the $D$--dimensional spacetime and we have  collectively denoted with
$y^i$ all the dynamical fields.

The variation $\delta_X  \L$ of the
Lagrangian density with respect to a vector field $X=X^i{\partial\over
\partial y^i}=\delta y^i{\partial\over \partial y^i}$ can be generally
written, through a well known  integration by parts procedure, as follows:
\begin{equation}\delta_X \L(j^{k} y)=e_i(j^{2k} y)X^i +d_\l
F^\l(j^{2k-1} y,j^{k-1}X)\label{FVFuno}
\end{equation} 
or, in terms of differential forms, as
\begin{equation}\delta_X L=e(L,X)+d F(L,X)\label{FVF}
\end{equation} 
Here the  $D$--form
\be
e(L,X)=e_i(j^{2k} y)X^i\,ds
\ee
and the $(D-1)$--form
\be
F(L,X)= F^\l(j^{2k-1}
y,j^{k-1}X)\,ds_\l\,, \qquad ds_\l=i_\l\rfloor\, ds
\ee
are called, respectively, \emph{the Euler--Lagrange  form} and
\emph{the Poincar\`e--Cartan  form}.
Obviously field solutions  are those field configurations 
$ y^i=\s^i(x)$  which  satisfy the
Euler--Lagrange equations: 
\begin{equation}
\left.e_i(j^{2k}y)\right\vert_{y=\s(x)}=0\label{25}
\end{equation} 
Let us now suppose that $\Xi$ is an infinitesimal generator of
symmetries, namely a projectable vector field $\Xi=\xi^\mu(x)
{\partial\over\partial x^\mu}+\Xi^i(x,y){\partial\over\partial y^i}$
which leaves field equations invariant:
\be
\delta_\Xi e_i(j^{2k}y)=0
\ee
The contraction of the  field equations \form{25} with the Lie derivatives
$\pounds_\Xi y^i$ of the field originates, apart for a sign, the
so--called \emph{variational} Lagrangian 
\be
L'(L, \Xi) := -e_i(j^{2k} y)\pounds_\Xi y^i\,ds
\ee
which depends on the dynamical fields  $y^i$ and the components
$\xi^\mu$ and $\Xi^i$ of the vector field $\Xi$; it is clearly vanishing
along solutions. According to the first variation formula
\form{FVF}, the  variation of  the variational Lagrangian $L'$
   yields:\footnote{
We remark that,  while it is not restrictive  to set $\delta \xi^\mu=0$
as the components $\xi^\mu$ do not depend on the dynamical fields, 
 we shall instead allow 
$\delta
\Xi^{i}\neq 0$.}
\begin{equation}
\delta_X L'=e(L',X)+d F(L',X)\label{FVF2}
\end{equation} 
Moreover, since we have assumed that $\Xi$ is an infinitesimal generator
of symmetries it turns out that the variational Lagrangian is a pure
divergence and therefore it is variationally trivial; see \cite{EM}. This
means that
$e(L',X)=0$ and, consequently, \form{FVF2} reduces to the conservation law
\be
d F(L',X)=\delta_X L'\simeq 0\label{31}
\ee
where $\simeq$ denotes equality on--shell. Moreover the
Poincar\`e--Cartan morphism $F(L',X)$ in \form{31} is linear in the Lie
derivatives $\pounds_\Xi y$ and their derivatives up to a fixed order.
Since in  all physically relevant  theories (e.g. in all gauge
natural theories; see \cite{libroFF,Remarks,kolar}) the Lie derivatives  
of the fields with respect to a vector field can be written as  linear
combinations   in the components of the vector field itself together with 
their (covariant) derivatives, it turns out that also the
Poincar\`e--Cartan morphism
$F(L',X)$  is a linear combination of the coefficients $\xi^\mu$ and
$\Xi^i$ of the vector field $\Xi$ together with  their derivatives up to a
finite order which depends on the particular theory we are dealing with.
This allows us to implement the so--called Spencer cohomology
\cite{libroFF,Remarks,Robutti,Spencer}. 
Namely, the Poincar\`e--Cartan
morphism
$F(L',X)$, through repeated  covariant integrations by parts with respect
to the components of $\Xi$, can be alghoritmically rewritten as
\cite{EM}: 
\be
F(L',X)=\tilde F(L',X)+d\U (L',X)\simeq d\U (L',X)\label{32}
\ee
where the term $\tilde F(L',X)$ turns out to be proportional to the field
equations together with  their derivatives and is thence vanishing
on--shell, while the
$(D-2)$ form
$\U (L',X)$ is the  (covariant) \emph{potential}. 
The potential  depends in turn  on the dynamical fields
$y^i$, on the components $(\xi^\mu,\Xi^i)$ of the infinitesimal generator
of symmetries and on the  components of the variational field $X$
together with  their derivatives up to a suitable order. Given a field
configuration
$y=\sigma(x)$ and a $(D-1)$--dimensional  region $\Sigma$ of spacetime
with boundary
$\partial \Sigma$, the integral
\ba
\delta_X Q_\Sigma(\sigma,\Xi):=\int_\Sigma F(L',X)&=&\int_\Sigma \tilde
F(L',X)+\int_{\partial \Sigma}\U (L',X)\\ &\simeq&\int_{\partial \Sigma}\U
(L',X)\label{34}
\ea
is assumed to be \emph{the variation} $\delta_X Q$ of the charge $Q$
along a one--parameter family of field configuration described by $X$ 
and relative to the symmetry originated by $\Xi$. If $y=\sigma(x)$ is a
solution of field equations and $X$ is a solution of the linearized field
equations (i.e. $X$ is tangent to the space of solutions) the quantity
$\delta_X Q_\Sigma(\sigma,\Xi)\simeq\int_{\partial \Sigma}\U (L',X)$
describes the variation of the charge $Q$ moving from $\sigma$ to nearby
solutions along a path in the space of solutions determined by
$X$. Moreover if the region $\Sigma$ is a (portion of) a Cauchy
hypersurface with a  boundary $\partial \Sigma$ and, in addition, the
spacetime projection
$\xi$  of the vector field
$\Xi$  is transverse to $\Sigma$,  we agree to identify  
$\delta_X Q_\Sigma$ with the variation of the energy  $
E(\partial \Sigma,\xi,X)$ enclosed inside $\partial
\Sigma$. We stress that, once we fix $\partial \Sigma$, there exist many
``energies'', depending both on the symmetry vector field $\xi$ and on the
variational vector field $X$; \cite{HIO,BY,forth,kij}. The
former vector determines, via its flow parameter, how the hypersurface
$\Sigma$ evolves as the ``time'' flows. The vector field $X$   fixes
instead the boundary conditions on the dynamical fields (e.g. Dirichlet or
Newmann boundary conditions). Each choice of the pair $(\xi,X)$ gives rise
to a particular realization of  a physical system endowed thence with its
own  ``energy content''.

Summarizing, we obtained the variation of conserved quantities
starting from the equations of motion through a two--step procedure. The
first step is nothing but an integration by parts of the variational
Lagrangian with respect to the vector field $X$ (see \form{FVF2}). The
second step amounts to perform a second integration by parts of the 
Poincar\`e--Cartan morphism
$F(L',X)$ with respect to the vector field $\Xi$ (see \form{32}). 
We remark that, a priori, there exists no guarantee  that the resulting
quantity $\delta_X Q_\Sigma$ is in fact the variation  of a
functional $Q$ of the intrinsic  geometry  of the boundary $\partial
\Sigma$. This becomes true  only if appropriate  boundary conditions are
chosen; see \cite{Anco,Barnich,Booth,BY,Nester,forth,HawHun,kij,RT}. 

The
viability of such a procedure has been tested in \cite{EM} in the realm of
natural and gauge natural theories of gravitations. The physically
expected values (or better, the commonly accepted expressions) for
conserved quantities, in particular for the energy, have been there
reproduced.\footnote{In
the application \cite{EM,forth} of the formalism to stardard
gravity we found that the Brown--York quasilocal energy \cite{BY}
 can be obtained when $\xi=\partial_t
$ is the vector generating an ADM foliation of spacetime and  $X$
corresponds to Dirichlet boundary conditions of the induced boundary
three--metric.} Encouraged by those positive results we aim now to
apply the formalism to Chern--Simons and Lovelock theories.\\

The variational Lagrangian $L'(L_1,\Xi)$ for the Chern--Simons
theory is easily obtained from the equations of motion \form{eq1} :
\be
L'(L_1,\Xi)=-{ \kappa}   
\epsilon_{A_1B_1
\dots A_{n+1}B_{n+1}} F^{A_1 B_1}\wedge\dots\wedge 
F^{A_nB_n} \pounds_\Xi A^{ A_{n+1}B_{n+1}}\label{LprimoCS}
\ee 
Taking into account that the Lie derivative \form{18} can be conveniently
rewritten in a manifestly covariant form as:
\be
\pounds_\Xi A^{AB}_\lambda=\xi^\rho F^{AB}_{\rho\lambda}+
{\stackrel{(A)}D}_\lambda \, \Xi^{AB}_{_{(V)}}\qquad
(\,\Xi^{AB}_{_{(V)}}=\Xi^{AB}+ A^{AB}_\sigma \xi^\sigma)\label{36}
\ee
the variation of \form{LprimoCS} turns out to be (first step):
\be
\delta_X L'(L_1,\Xi)=d_\alpha \F^\alpha (L'(L_1,\Xi),X)
\ee
where
\ba
\F^\alpha (L'(L_1,\Xi),X)&=&-{\kappa\over 2^n}
\epsilon^{\lambda\mu\nu\mu_1\nu_1\dots \mu_{n-1}\nu_{n-1}}
\epsilon_{ABCD A_1B_1
\dots A_{n-1}B_{n-1}}\nonumber\\
 &&\times F^{A_1 B_1}_{\mu_1\nu_1}\dots 
F^{A_{n-1}B_{n-1}}_{\mu_{n-1}\nu_{n-1}}\label{37}\\
&&\times\left\{2n \,\delta^\alpha_\mu
\delta A^{AB}_\nu\pounds_\Xi 
A^{CD}_\lambda+F^{AB}_{\mu\nu}\left(2\xi^{[\alpha}
\delta^{\gamma]}_\lambda
\delta
A^{CD}_\gamma+
\delta^\alpha_\lambda\delta\Xi^{CD}_{_{(V)}}\right)\right\}\nonumber
\ea
Inserting \form{36} into \form{37} and integrating by parts the term
containing the covariant derivative we obtain (second step):
\be
\F^\alpha (L'(L_1,\Xi),X)=\tilde\F^\alpha (L'(L_1,\Xi),X)+d_\beta
\U^{\alpha\beta}(L'(L_1,\Xi),X)
\ee
where
\be
\tilde\F^\alpha (L'(L_1,\Xi),X)=\delta \left\{
-{\kappa\over
2^n}\epsilon^{\alpha\mu_1\nu_1\dots\mu_n\nu_n}
\epsilon_{ABA_1B_1\dots A_nB_n}F^{A_1B_1}_{\mu_1\nu_1}\dots
F^{A_nB_n}_{\mu_n\nu_n}\Xi^{AB}_{_{(V)}}
\right\}
\ee
vanishes on--shell,  while 
\ba
\U^{\alpha\beta}(L'(L_1,\Xi),X)&=&{n\kappa \over 2^{n-1}} 
\epsilon^{\alpha\beta\rho\mu_1\nu_1\dots
\mu_{n-1}\nu_{n-1}}\epsilon_{ABCDA_1B_1\dots A_{n-1}B_{n-1}}
\nonumber\\
&&\times F^{A_1 B_1}_{\mu_1\nu_1}\dots 
F^{A_{n-1}B_{n-1}}_{\mu_{n-1}\nu_{n-1}}\; \delta
A^{AB}_\rho\Xi^{CD}_{_{(V)}}
\label{41A}
\ea
are the coefficients of the \emph{covariant} potential $(2n-1)$--form
$\U={1\over2}\U^{\alpha\beta} ds_{\alpha\beta}$ with
$ds_{\alpha\beta}=i_\beta\rfloor\, i_\alpha\rfloor \, ds$ (an expression
similar to \form{41A} was also obtained in 
\cite{SilvaCS} through  a different technique).  We remark
that the integration of the potential gives rise, 
 through expression
\form{34}, to the variation of the charge associated with the
infinitesimal symmetry $\Xi$. Notice that the potential can be 
rewritten  in a more compact form  as follows\footnote{
According to our notation 
\be
dx^{\mu_1}\wedge dx^{\nu_1}\wedge\dots\wedge dx^{\mu_{n-1}}\wedge
dx^{\nu_{n-1}}\wedge dx^\rho={1\over2}\epsilon^{\alpha\beta\rho\mu_1 \dots
\nu_{n-1}}\; ds_{\alpha\beta}
\ee}:
\ba
\U(L'(L_1,\Xi),X)&=&{n\kappa }\; 
\epsilon_{ABCDA_1B_1\dots A_{n-1}B_{n-1}}
\nonumber\\
&&\times F^{A_1 B_1}\wedge\dots \wedge
F^{A_{n-1}B_{n-1}}\wedge\delta
A^{AB}\Xi^{CD}_{_{(V)}}
\label{41}
\ea

Since the potential has been obtained directly from the
equations of motion and the equations of motion \form{eq1} are equivalent
to field equations \form{eqq2} we can immediately obtain the potential 
$\U(L'(L_2,\Xi),X)$ relative to the latter equations.
According to the schema \form{transition2} it is given by:
\ba
&&\U(L'(L_2,\Xi),X)={n\kappa\over l } 
\epsilon_{abca_1b_1\dots a_{n-1}b_{n-1}}
\nonumber\\
&&\times \left\{\hat R^{a_1 b_1}\wedge\dots \wedge
\hat R^{a_{n-1}b_{n-1}}\wedge (\Xi^{ab}_{_{(V)}}\;\delta
e^{c}+\Xi^c_{_{(V)}}\delta\omega^{ab})\right.
\label{xxx}\\
&&\left.+\hat R^{a_2 b_2}\wedge\dots 
\hat R^{a_{n-1}b_{n-1}}\wedge T^c\wedge\delta
\omega^{ab} \Xi^{a_1b_1}_{_{(V)}}\right\}
\nonumber
\ea
with
\be
\Xi^{a}_{_{(V)}}= e^a_\mu \xi^\mu,\qquad
\Xi^{ab}_{_{(V)}}=\Xi^{ab}+ \omega^{ab}_\mu\xi^\mu
\ee
The result \form{xxx} can be easily inferred by inserting the splittings
\form{AWE} and
\form{FWE} into \form{41} and  recalling the prescription $\Xi^a=0$
for the generators of symmetries. Alternatively, expression \form{xxx}
can be obtained,  in a clearly more involved way, from the variational
Lagrangian 
\ba
L'(L_2,\Xi),X)&=&- (\kappa /l)
\epsilon_{a a_1 b_1\dots a_n b_n} 
\hat{R}^{a_1b_1}\wedge\dots\wedge  \hat{R}^{a_nb_n}\wedge\pounds_\Xi
e^a\\
&&-(n \kappa/l)
 \epsilon_{a a_1 b_1\dots a_n b_n} 
 T^a\wedge \hat{R}^{a_1 b_1}\wedge\dots\wedge\hat{R}^{a_{n-1}b_{n-1}}
\wedge\pounds_\Xi\omega^{a_nb_n}\nonumber
\ea
(obtained from the contraction of the equations \form{eqq2} with the Lie
derivatives of fields with respect to the vector field \form{21}) via the
two--step procedure outlined above. 

Let us now turn the attention to the Lovelock--Chern--Simons theory
described through the field equations ${\delta L_3\over \delta
g_{\lambda\rho}}={\partial L_3\over \partial
g_{\lambda\rho}}$ ensuing from the Lagrangian
\form{LGB}. \footnote{
We remark that field equations are simply obtained  by varying  $L_3$
with respect to the metric alone since the variation of the curvature
terms  yields  a total derivative. This is the reason why field
equations are of second order only; \cite{lanczos,lov}.}
The two--step calculation  of the conserved quantities
starting  from the variational Lagrangian $L'(L_3,\xi)=-{\partial L_3\over \partial
g_{\lambda\rho}}\pounds_\xi g_{\lambda\rho}$ is obviously a rather
cumbersome task. Nevertheless we have outlined in the introduction that 
conserved quantities  can be calculated directly from \form{xxx} if we set
$T=0$ and if we are able to select  a suitable \emph{covariant} lift of
the spacetime vector field $\xi$ up to a vector field $\Xi$ into the
configuration bundle $Y$  of the Lagrangian $L_2$. Namely if we are able
to select  a suitable rule $\Xi^{ab}=\Phi^{ab}(\xi^\mu, d_\nu\xi^\mu,
\dots)$ which determines the coefficients of $\Xi$  in terms of
the components of $\xi$ and, obviously, in terms of the dynamical fields;
see diagram
\form{transition2}. As already explained the prescription which allows to
rule out the sought--for lift is the requirement that 
 the  variational Lagrangian
$L'(L_2,
\Xi)$  and the  variational Lagrangian
$L'(L_3,
\xi)$ are equivalent (so that conserved quantities calculated from them
are equivalent, too).

According to \cite{CS3d,Koslor,EM,Godina,vangoden}   the 
\emph{generalized Kosmann lift}
$\Xi=K(\xi)$, obtained by the rule 
\be
{}^{K}\!{\xi^{ab}}={}^{K}\!{\xi^{[ab]}}=e^{[a}_\alpha \,
e^{b]\lambda} d_\lambda
\xi^\alpha-\xi^\gamma e^{[a}_\mu d_\gamma e^{b]\mu},\label{Kosmann}
\ee
is the only one which restores the equality between the variational
Lagrangians. This property is easily proved. Indeed, with the expression
\form{Kosmann} we have 
\be
\pounds_{K(\xi)}  e^a_{\mu}={1\over 2} e^{a\nu} \pounds_\xi
g_{\mu\nu}\label{56}
\ee
so that
\ba
\left.L'(L_2,K(\xi))\right\vert_{T=0}&=&
-{\delta L_2\over \delta
e^a_{\mu} }\pounds_{K(\xi)} e^a_{\mu}
=
-{\partial L_2\over \partial
e^a_{\mu} }\pounds_{K(\xi)} e^a_{\mu}\nonumber\\
&=&
-{1\over 2}{\partial L_2\over
\partial e^a_{\mu} } e^{a\nu} \pounds_\xi
g_{\mu\nu}=-{\partial L_3\over \partial
g_{\mu\nu}} \pounds_\xi g_{\mu\nu}\nonumber\\
&=&-{\delta L_3\over \delta
g_{\mu\nu}} \pounds_\xi g_{\mu\nu}=L'(L_3,\xi)\nonumber
\ea
where we have taken into account that, if $T=0$, then
$\left.L_2(e,j^1 \omega)\right\vert_{T=0}=L_3(g(e))$ and ${1\over
2}{\partial L_2\over
\partial e^a_{\mu} } e^{a\nu} ={\partial L_3\over \partial
g_{\mu\nu}}$.

We also stress  that the lift \form{Kosmann}  
is globally well defined. Indeed, the Kosmann lift $K(\xi)$
of a spacetime vector field $\xi$
 transforms tensorially (as it can be inferred through a direct
inspection) and can be therefore globally defined. Namely, all  local
expressions
\form{Kosmann} can be patched together  to define a unique global vector
field; see \cite{Koslor,Godina,vangoden}.

It follows that the potential for the Lovelock--Chern--Simons theory can
be obtained directly from \form{xxx} by setting $T=0$ and taking into
account
 the prescription
\form{Kosmann}, i.e.:
\be
\cases{
\U(L'(L_3,\xi),X)=
{n\kappa\over l } 
\epsilon_{abca_1b_1\dots a_{n-1}b_{n-1}}(
 {}^{K}\!\xi^{ab}_{_{(V)}}\;\delta e^{c}+{}^{K}\!\xi^c_{_{(V)}}
\delta\omega^{ab})
\cr
\phantom{\U(L'(L_3,\xi),X)=}
\wedge \hat R^{a_1 b_1}\wedge\dots 
\wedge\hat R^{a_{n-1}b_{n-1}}
\cr
\cr
{}^{K}\!{\xi^{a}_{_{(V)}}}= e^a_\mu \xi^\mu,\qquad
{}^{K}\!{\xi^{ab}_{_{(V)}}}=e^{[a}_\alpha e^{b]\beta}\nabla_\beta
\xi^\alpha
}
\label{masterino}
\ee
According to \form{34} the variation of conserved quantity (and
therefore the variation of energy) is then obtained from the above
expression after a suitable integration on a  spacetime region.

Let us remark that it is also possible to separate into \form{masterino}
the contributions ensuing from each Lagrangian $\bar\L_p$ of \form{LGB}.
Indeed,   if we take  the definition
\form{12} into account  expression \form{masterino} can be conveniently
rewritten as:
\be
\U(L'(L_3,\xi),X)=\kappa \sum_{p=0}^{n}\bar c_p
\U_{p}\label{summa}
\ee
where 
\ba
\U_p&=&
{p\over (2n-2p)! }
\epsilon_{abc a_1b_1\dots a_{n-1}b_{n-1}} \left\{(
e^{a}_\alpha e^{b\beta}\nabla_\beta
\xi^\alpha)\delta e^{c} +(e^c_\mu \xi^\mu)
\delta\omega^{ab}\right\}\nonumber\\ &&
\wedge R^{a_1 b_1}\wedge\dots 
\wedge R^{a_{p-1}b_{p-1}}\wedge
e^{a_{p}}\wedge e^{b_{p}}\wedge\dots\wedge
e^{a_{n-1}}\wedge e^{b_{n-1}}\;
\label{masterino2}
\ea
and $\bar c_p$ are given by \form{CP}.
A  direct comparison between \form{summa} and \form{LGB} leads to our
master formula
\be
\U(\bar \L_p)=\U_{p}
\ee
for the \emph{covariant} potential relative to the term $\bar \L_p$.
Therefore we are now able  to calculate the potential for a generic
Lovelock Lagrangian
\be
L_{Lov}=\sum_{p=0}^n  \alpha_p\bar \L_p(j^2g)\label{Lov}
\ee
with \emph{completely} arbitrary constant coefficients $ \alpha_p$. It
turns out to be:
\be
\U_{Lov}=\sum_{p=0}^{n}
 \alpha_p\U_{p}\label{master}
\ee
In particular $\bar\L_0=\sqrt{g}$ gives no contribution to the potential,
while the contribution ensuing from the Hilbert term 
$\bar\L_1=\sqrt{g}\; R$ is: 
\be
\U^{\alpha\beta}_1=\sqrt{g}\;\delta^{\alpha\beta\rho}_{\mu\nu\sigma}
(\nabla^\nu\xi^\mu e^\sigma_c \delta e^c_\rho +\xi^\sigma
e^\mu_a e^\nu_b \delta \omega^{ab}_\rho)\label{ULH}  
\ee

%\end{document}

Formula \form{ULH}   reproduces
exactly the formula for the  variation of  conserved quantities   found in
\cite{Booth,BY,forth,kij}  for General Relativity. This property was
proven explicitly in \cite{CS3d} when
the   spacetime dimension is three; but the same calculation can be
carried on, step by step, in any higher spacetime dimension.
 We also point out the  remarkable simple expression
\form{masterino2} and \form{master} for the potential in Lovelock
gravity.  In the sequel we shall test its
viability, i.e. we shall analyse whether  formula \form{master} allows to
reproduce physically expected quantities in specific applications.

\begin{Remark}
\label{Remark}{\rm
Let us consider again formula \form{41}. It is straightforward to verify
that, on--shell, the potential satisfies:
\be
d\,\U(L'(L_1,\Xi),X)= \omega(L_1, X,\pounds_\Xi
A)\label{522} 
\ee
where the $(D-1)$--form: 
\ba
\omega(L_1, X,\pounds_\Xi
A)&=&{n\kappa } 
\epsilon_{ABCDA_1B_1\dots A_{n-1}B_{n-1}}
\nonumber\\
&&\times F^{A_1 B_1}\wedge \dots \wedge
F^{A_{n-1}B_{n-1}}\wedge\pounds_\Xi 
A^{AB}\wedge\delta_X A^{CD}
\ea
is the symplectic  current relative to $L_1$; see \cite{SilvaCS}.
This    remarkable property ensures  that we can make the
identification 
\be
\U(L'(L_1,\Xi),X)=\delta_X \V(L_1,\Xi)-B(L_1,\Xi,X)
\ee
where $\V(L_1,\Xi)$ is the Noether superpotential, while $B(L_1,\Xi,X)$ is
the so--called Regge--Teitelboim covariant correction term, namely the
term which has to be added to the variation of the Hamiltonian to have a
well--posed variational problem (or, in other words, it is the factor 
which exactly cancels  the boundary term coming from the variation of the
Hamiltonian; see \cite{CADM, Wald95,Silva, RT}). This property, a
posteriori, endows the definition we suggested for the variation of the
energy, i.e:
\be
\delta_X E_\Sigma(\Xi)= \int_{\partial
\Sigma}\U(L'(L_1,\Xi),X)\simeq \int_\Sigma 
\omega(L_1, X,\pounds_\Xi
A) \label{5444}
\ee
with a well-defined formal meaning since it is coherent with a
(Hamiltonian) symplectic  description of field theories.

Moreover,
 if $\Xi$ is a Killing vector  for the solution, i.e.
$\pounds_\Xi A=0$,  the symplectic form $\omega$ is
vanishing, the p"otential $\U$ becomes a closed form  and
formula \form{5444} is the conservation law for energy; see
\cite{CS3d,Silva,kij}.\\

Inserting \form{AWE} and \form{FWE} into \form{522} and setting  $T=0$ 
and
$\Xi=K(\xi)$,   we easily obtain the
conservation law for Lovelock--Chern--Simons theory: 
\be
d\,\U(L'(L_3,K(\xi)),X)= \omega(L_3,
X,\pounds_{K(\xi)}
\omega, \pounds_{K(\xi)}
e)\label{52222} 
\ee
where the symplectic current \cite{TZ} is now given by:
\ba
\omega(L_3, X,\pounds_\Xi
\omega, \pounds_\Xi
e)&=&{n\kappa \over l}\; 
\epsilon_{abca_1b_1\dots a_{n-1}b_{n-1}}
 \hat R^{a_1 b_1}\wedge\dots \wedge
\hat R^{a_{n-1}b_{n-1}}\nonumber\\
&&\wedge (\pounds_{K(\xi)} e^a
\wedge \delta \omega^{bc}-
\delta e^a\wedge \pounds_{K(\xi)} \omega^{bc})
\ea
and is vanishing whenever $K(\xi)$ is a Killing vector for the dynamical
fields $(e^a,\omega^{bc})$.\footnote{We remark that the Lie derivative of
the vielbein with respect to the Kosmann lift  is given by \form{56},
while the Lie derivative of
the spin connection turns out to be \cite{EM}:
\be
\pounds_{K(\xi)} \omega^{ab}_\mu=e^{[a}_\lambda\, e^{b]\gamma}\pounds_\xi
\gamma^\lambda{}_{\gamma\mu}\label{561}
\ee
where $\gamma^\lambda{}_{\gamma\mu}$ are the Christoffel symbols of the
metric $g_{\mu\nu}=\eta_{ab} e^a_\mu e^b_\nu$. }
\CVD}\end{Remark}

\begin{Remark}{\rm
The electromagnetic field can be included into the theory simply 
by minimal coupling the electromagnetic field to the
Lovelock--Chern--Simons  Lagrangian \form{LGB}, i.e. 
\be
L_{Lov}^{em}=L_3(j^2 g)+L_{em}(g, F)\label{61}
\ee
where $L_{em}(g, F)$ is 
the Maxwell Lagrangian in $(2n+1)$ dimensions: 
\be
L_{em}(g, F)=-{1\over  4\Omega_{D-2}}\sqrt{g}
F_{\mu\nu}F_{\alpha\beta}g^{\mu\alpha} g^{\nu\beta}
\ee
The potential  relative to the pure
electromagnetic part can be  calculated, through the two--step
 procedure
outlined in section \ref{Conserved Quantities}, starting  from the
variational Lagrangian
\be
L'(L_{em},(\xi,\chi))=-{\delta L_{em}\over \delta A_\mu}
\pounds_{(\xi,\chi)} A_\mu- {\delta L_{em}\over \delta g_{\mu\nu}}
\pounds_\xi g_{\mu\nu}
\ee
where
\be
\cases{
\pounds_{(\xi,\chi)} A_\mu=\xi^\rho d_{\rho}A_\mu +d_\mu \xi^\rho A_\rho
+d_\mu
\chi\cr
\cr
\pounds_\xi g_{\mu\nu}=\nabla_\mu\xi_\nu +\nabla_\nu\xi_\mu }\label{604}
\ee
We shall here skip the details of the calculation since they are just a
mere and straightforward application of the formalism. The result turns
out to be:
\ba
\U^{\alpha\beta}(L'(L_{em},(\xi,\chi)),X)&=&-{1\over  \Omega_{D-2}}\delta
(\sqrt{g} F^{\alpha\beta}) \; (\chi+ A_\rho
\xi^\rho)\nonumber \\
&&-{\sqrt{g}\over  \Omega_{D-2}}(F^{\alpha\beta}\xi^\rho +2
\xi^{[\alpha}F^{\beta]\rho})\;\delta A_\rho\label{66}
\ea
It then follows the the potential for the Lagrangian \form{61} is
obtained from \form{masterino} and \form{66}, i.e.
\be
\U^{\alpha\beta}(L'(L_{Lov}^{em},{(\xi,\chi))},X)=\U^{\alpha\beta}
(L'(L_3,\xi),X)+
\U^{\alpha\beta}(L'(L_{em},{(\xi,\chi)}),X)\label{67}
\ee
\CVD}\end{Remark}

%%%%%%%%%%%%%%%%%%%%%%%%%%%%%%%%%%%%%%%%%%%%%%%%%%%%%%%%%%%%%%%%%%%%%%%%%%%%%%%%%%%%%%%%%%%%%%%%%%%%%%%%%%%%%%%%%%%%%%%

\section{The $D=5$  Spherically Symmetric Charged Solution}
\label{The spherically symmetric charged solution}
To check the reliability of the formula \form{67}  we test it on an exact
solution and we show that it allows to recover the correct expression for
the first law of black holes mechanics when stationary black holes
solutions  for the
$5$-dimensional Lovelock--Maxwell gravity are considered 
\cite{Bana5d,Bana5d1,boul,BHS,deha}.\\

Let us consider a five dimensional solution of the field
equations ensuing from the $D=5$ Lagrangian \form{61}.  The gravitational
contribution to the total  gravitation--electromagnetic potential  
\form{67} can be read out from formula \form{summa} with $n=2$. It is given
by:
\be
\U^{\alpha\beta}(L'(L_3,\xi),X)=
\U^{\alpha\beta}(L'( L_H,\xi),X)+\U^{\alpha\beta}(L'(
L_{GB},\xi),X)
\label{7b5}
\ee
where
\be
\U^{\alpha\beta}(L'( L_H,\xi),X)={ 4\kappa\over 
l^3}\sqrt{g}\;\delta^{\alpha\beta\rho}_{\mu\nu\sigma}
(\nabla^\nu\xi^\mu e^\sigma_c \delta e^c_\rho +\xi^\sigma
e^\mu_a e^\nu_b \delta \omega^{ab}_\rho)
\ee
and:
\be
\U^{\alpha\beta}(L'(L_{GB},\xi),X)={ 1\over 
12\pi^2}\epsilon^{\alpha\beta\rho\sigma\nu}\epsilon_{abcde} \;
{R}_{\rho\sigma}^{ab}\Big(e^e_\mu \xi^\mu \delta \omega^{cd}_\nu +e^d_\mu
e^{e\gamma}\nabla_\gamma\xi^\mu \delta e^c_\nu\Big)\label{7b7}
\ee
are the contributions ensuing, respectively, from  the Hilbert Lagrangian
$\bar
\L_1$  and the Gauss--Bonnet Lagrangian $\bar \L_2$. 
The electromagnetic
contribution to the total  potential  
\form{67} is instead given by   formula \form{66}:
\ba
\U^{\alpha\beta}(L'(L_{em},(\xi,\chi)),X)&=&-{1\over  2\pi^2}\delta
(\sqrt{g} F^{\alpha\beta}) \; (\chi+ A_\rho
\xi^\rho)\nonumber \\
&&-{\sqrt{g}\over  2\pi^2}(F^{\alpha\beta}\xi^\rho +2
\xi^{[\alpha}F^{\beta]\rho})\;\delta A_\rho\label{6666}
\ea
We remark that the area $\Omega_3$ of the three dimensional unit sphere
is $12\pi^2$ so that, from \form{costanteacc}, we infer that  the coupling
constant
$\kappa$ turns out to be
$\kappa= l/12\pi^2$.\\

The spherically symmetric charged black hole solution is; 
see \cite{Bana5d}:
\be
\cases{
ds^2=-f^2 (r) d t^2+ \frac{1}{f^2 (r)} d r^2 +r^2 d \Omega_3\cr
\cr
f^2 (r)=1+ \frac{r^2}{l^2} -\sqrt{  M+1-{\frac{{Q}^2}{4
r^2}}}\cr
\cr
A=\varphi dt, \qquad \varphi=\frac{Q}{2 r^2}}
\label{68}\ee
where $d \Omega_3$ is the standard metric on the three--dimensional  unit
sphere.
The event horizon is located at $r=r_+$, where $r_+$ is the positive
solution of
$f^2 (r_+)=0$; see \cite{Bana5d}. If we now consider the vector field
${(\xi,\chi)}=(\partial_t,\chi_0)$, with $\chi_0=constant$, and we
integrate expression \form{67} on a three--sphere $S$ of constant $r$ and
$t$ (with $r>r_+$) we obtain 
\be
\delta_X Q_{3}(S,\xi)=\delta M -{Q\delta Q\over 2 r^2}\label{800}
\ee
from the gravitational part \form{7b5}; and 
\be
\delta_X Q_{em}(S,\xi,\chi_0)={Q\delta Q\over 2 r^2}+\chi_0 \delta Q
\label{801}
\ee
from the electromagnetic part \form{6666}. The variation of the total
\emph{charge} is therefore obtained by their sum:
\be
\delta_X Q_{Lov}^{em}(S,\xi,\chi)=\delta M+\chi_0 \delta Q\label{771}
\ee
and reproduces  the espected physical values.
We remark that even though \form{800} and \form{801}  depend explicitly 
on
the radius of the sphere, their   sum \form{771} does not depend on it.
This property follows from the consideration we made in  Remark
\ref{Remark}. Indeed from
\form{604}  we can easily infer that the vector field
$(\xi,\chi_0)$ is a Killing vector field  for the solution \form{68} and
consequently the potential
\form{67} is a closed three--form. Therefore:
\be
\delta_X Q_{Lov}^{em}(S,\xi,\chi_0)=\delta_X
Q_{Lov}^{em}(S',\xi,\chi_0)\label{72}
\ee 
whenever $S$ and $S'$
are homologous surfaces. 

In particular,  we can identify $\chi_0$ with
(minus) the electric potential on the horizon
$\chi_0=-\varphi(r_+)=-\frac{Q}{2 r_+^2}$ (the thermodynamical 
parameter  conjugated to the charge). In this
case formula
\form{771} becomes the first law of black hole thermodynamics: 
\be
\delta M-\varphi(r_+) \delta Q=T \delta S\label{73}
\ee
where:
\be
T=\frac{1}{4 \pi} \left[   \frac{2 r_+}{l^2} -{{Q}^2 \, l^2\over 4
r_+^3(l^2+r_+^2)}  
    \right]
\ee
is the temperature $T=\frac{1}{4 \pi}\left.{d\,\, f^2(r)\over
dr}\right\vert_{r=r_+}
$of the black hole, while 
\be
S=8 \pi \left[ r_{_+}  + {r_{_+}^3 \over 3l^2}\right]
\label{entr}
\ee
 is its entropy; see
 \cite{Bana5d}. In obtaining this result,  in the right hand side of
\form{73}
 we  have taken into account that the mass and the
charge of the solution are related to the radius of the horizon $r_+$ via
the relation
$f^2 (r_+)=0$.

%%%%%%%%%%%%%%%%%%%%%%%%%%%%%%%%%%%%%%%%%%%%%%%%%%%%%%%%%%%%%%%%%%%%%%%
\section{The Entropy in Lovelock Gravity}
\label{entropia} 
As it is manifest from \form{entr} the entropy  for the spherically
symmetric charged solution differs from the area law of Einstein
gravity. This is a well--known result. It was indeed demonstrated in
\cite{T.Jacobson}  that  in Lovelock
gravity   each higher
derivative term gives a contribution to the entropy in its own.
We shall now analyse to what extent formulae 
\form{masterino2} and \form{master} are in agreement with the results
already existing in literature.\\

Let us consider the Lovelock Lagrangian \form{Lov} and 
 let us consider  a stationary  black hole solution of its field
equations, admitting a Killing horizon $\H$ and  a bifurcation surface
$H$; see
\cite{Wald,T.Jacobson, Towsend,Waldbook}. Let us denote by $\xi$ the
unique  Killing vector field  which is null on the horizon and vanishes
on $H$, i.e. the vector field which describes the evolution of a system
of ROTORS (co--rotating observers; see
\cite{visser}). We remark that on the bifurcation surface it holds true
that:  
\be
\nabla_\mu \xi_\nu= -2\kappa_H \beta_{\mu\nu}\label{78}
\ee
 where $\kappa_H$  is
the surface gravity and
 $\beta_{\mu\nu}$ denotes the binormal to $H$.

Let us then consider  a spacelike  hypersurface $\Sigma$ which 
extends from spatial infinity  to the horizon and
intersects
the horizon in the bifurcation surface.  Since $\xi$ is transverse to $D$
the variation of energy is identified  with the the integral $\int_S
\U(\xi,X)$ of the potential $\U$  on a $(2n-1)$--dimensional  surface $S$
enclosing the singularity. But  
the potential is a closed form since $\xi$ is  a Killing vector (see
Remark \ref{Remark}). Therefore the integral $\int_S
\U(\xi,X)$ is equal to $\int_{S'}
\U(\xi,X)$ for any  surface $S'$ homologous to $S$. In particular we can
identify $S'$ with the bifurcation surface  $H$
and   calculate  the variation of the energy on it:
\be
\delta E(\xi)=\int_H \frac{1}{2} \U^{\alpha\beta}_{Lov}
ds_{\alpha\beta}=\sum_{p=0}^{n}\alpha_p\int_H \frac{1}{2}
\U^{\alpha\beta}_p ds_{\alpha\beta}\label{79}
\ee
where the potentials $\U^{\alpha\beta}_p$ are given by 
\form{masterino2}. As  the vector $\xi$ vanishes on $H$ only terms
containing derivatives of
$\xi$ survive into
\form{79}, i.e.
\ba
 \left.\U^{\alpha\beta}_p\right\vert_H& =&
{p\over (2n-2p)! 2^{p-1}}
\epsilon^{\alpha\beta\rho\mu_1\nu_1\dots
\mu_{n-1}\nu_{n-1}}\epsilon_{abca_1b_1\dots a_{n-1}b_{n-1}}\nonumber \\
&& \times\left\{ R^{a_1 b_1}_{\mu_1\nu_1}\dots 
 R^{a_{p-1}b_{p-1}}_{\mu_{p-1}\nu_{p-1}}\,
e^{a_{p}}_{\mu_{p}}e^{b_{p}}_{\nu_{p}}\dots
e^{a_{n-1}}_{\mu_{n-1}}e^{b_{n-1}}_{\nu_{n-1}}\;\right\}\nonumber \\
&&\times\;e^a_\gamma
e^{b\sigma}\nabla_\sigma
\xi^\gamma \delta e^c_\rho\label{908}
\ea
Taking into account the property \form{78} together with the integration
rule  
\be
\int_S \frac{1}{2} \sqrt{g} 
\; \V^{[\alpha\beta]} ds_{\alpha\beta}=\int_S \sqrt{\tilde g}\, d^{D-2}x
\;\V^{[\alpha\beta]}\beta_{\alpha\beta}
\ee
 (which holds true for any $({D-2})$--form 
$\V=\frac{1}{2} \sqrt{g} 
\; \V^{[\alpha\beta]} ds_{\alpha\beta}$ and any $({D-2})$--region $S$ with
induced metric $\tilde g$ and binormal
$\beta_{\alpha\beta}=n_{[\alpha} u_{\beta]}$, where $n_\alpha$ and
$u_\beta$ are the outward pointing  unit normals to $S$) we obtain:
\ba
&&\int_H \frac{1}{2}
\U^{\alpha\beta}_p ds_{\alpha\beta}={2\;p\;\kappa_H\over (2n-2p)!
2^{p-1}}\int_H
\sqrt{\tilde g} d^{D-2}x\label{RRRR}\\
 &&
\times\tilde
\epsilon^{\rho\mu_1\nu_1\dots
\mu_{n-1}\nu_{n-1}}
\tilde\epsilon_{\lambda\alpha_1\beta_1\dots\alpha_{p-1}\beta_{p-1}
\mu_{p}\nu_{p}\dots\nu_{n-1}}
 R^{\alpha_1 \beta_1}_{\mu_1\nu_1}\dots 
 R^{\alpha_{p-1}\beta_{p-1}}_{\mu_{p-1}\nu_{p-1}}\;e^\lambda_c\delta
e^c_\rho
\nonumber
\ea
where $\tilde\epsilon^{\mu_1\dots\mu_{2n-1}}=
\epsilon^{\mu\nu\mu_1\dots\mu_{2n-1}}\beta_{\mu\nu}$ is the
$(2n-1)$--dimensional Levi--Civita  totally skew--symmetric  tensor
density (from now on a ``tilde'' over a quantity  will always denote the
projection of the quantity itself onto the surface $H$). Notice that in
the above formula  the Riemann tensors are fully projected onto the
surface $H$. Moreover, since $H$ is a bifurcation surface, its extrinsic
curvature vanishes and (via the Gauss--Codazzi equations
\cite{Gravitation, Waldbook}) we can identify the projected Riemann
tensor with the 
$(2n-1)$--dimensional Riemann tensor $\tilde R$ built out of the induced 
$(2n-1)$--metric $\tilde g_{\mu\nu}$, see
\cite{T.Jacobson}. Contracting  
the product of  the Levi--Civita tensor densities
in \form{RRRR} and rewriting  them as a product of  $(2n-1)$--dimensional
Kronecker deltas
$\tilde \delta^\mu_\nu$ we finally obtain:
\be
\delta E_p(\xi):=T\delta S_p={p\;\kappa_H\over 2^{p-2}}\int_H 
\sqrt{\tilde g}\, d^{D-2}x\;
\tilde\delta^{\rho\mu_1\nu_1\dots
\mu_{p-1}\nu_{p-1}}_{\lambda\alpha_1\beta_1\dots
\alpha_{p-1}\beta_{p-1}} \tilde R^{\alpha_1 \beta_1}_{\mu_1\nu_1}\dots 
\tilde R^{\alpha_{p-1}\beta_{p-1}}_{\mu_{p-1}\nu_{p-1}}\;\tilde
e^\lambda_c\delta
\tilde e^c_\rho\label{excr}
\ee
where the temperature $T$ is given by $T= {\kappa_H\over 2\pi}$.
As recognized  in
\cite{T.Jacobson}
the above expression can be formally integrated since:
\be
\sqrt{\tilde g}\; 
\tilde\delta^{\rho\mu_1\nu_1\dots
\mu_{p-1}\nu_{p-1}}_{\lambda\alpha_1\beta_1\dots
\alpha_{p-1}\beta_{p-1}} \tilde R^{\alpha_1 \beta_1}_{\mu_1\nu_1}\dots 
\tilde R^{\alpha_{p-1}\beta_{p-1}}_{\mu_{p-1}\nu_{p-1}}\;\tilde
e^\lambda_c\delta
\tilde e^c_\rho=2^{p-1} \delta \bar \L_{(p-1)}
\ee
Extracting  the temperature from \form{excr} we thence 
obtain:
\be
S_p=4\pi \;p\int_H \bar \L_{(p-1)}(\tilde g)
\ee
namely, the  contribution to the entropy ensuing from the  term $\bar
\L_{p}(g)$ of the Lovelock Lagrangian is obtained from the 
 Lagrangian  $\bar\L_{(p-1)}(\tilde g)$ defined with respect
to the induced $(2n-1)$--metric $\tilde g_{\mu\nu}$. Therefore \form{79}
corresponds to the first law of  black holes thermodynamics\footnote{
We remark that, according to our choice of the vector $\xi$, the variation
$\delta E$ of energy turns out to be $\delta
E=\delta M-\Omega^{(i)} J_{(i)}$ where: $M$ is the mass of the black
hole, $J_{(i)}$, with  $i=1\dots n$, are the angular momenta associated 
with the commuting Killing vectors generating spatial  rotations and
$\Omega^{(i)}$ are the horizon angular velocities; see
\cite{deha,T.Jacobson}.}:
\be
\delta E(\xi)= T \delta S
\ee
where 
\be
S=4\pi\sum_{p=1}^{n}p\,\alpha_p\,\int_H\bar \L_{(p-1)}(\tilde g)
\ee
This result 
agrees  exactly with the result obtained by  Wald within a Lagrangian
approach
\cite{Wald} or by  Jacobson-Myers within a  Hamiltonian method
\cite{T.Jacobson}. 

We stress again
that \form{79} is the integral of a closed form (since $\xi$ is a Killing
vector) and therefore the  integral on $H$ equals the integral on any
other surface $S$ homologous to
$H$ itself. Nevertheless on any other  surface other than  $H$ the formal
calculation of $\delta E(\xi)=\int_S U(\xi,X)$ becomes rather more
involved since also  the terms linear in
$\xi$  have to be retained  and the simplifications related to the
particular geometry of the bifurcation surface do not hold true any
longer. Despite of this, even though  the formal calculation above
requires the presence of the bifurcation surface, on the contrary, 
as we did already remark in \cite{Remarks,BTZ}, its
existence  become a unessential requirement in specific applications. In
the charged spherical symmetric solution we have dealt with in section
\ref{The spherically symmetric charged solution}, e.g., 
 the bifurcation surface is out of the domain of coordinates (and its
presence would be revealed only in the maximal analytic  extension
with Kruskal--Szekeres--like coordinates;
\cite{HEllis,Wald,Racz,Towsend}).  However the first law \form{73} of
black hole mechanics has been established performing calculation on
\emph{any} surface enclosing the singularity. This noteworthy property, we
stress again,  follows from the cohomological properties the  potential
inherits in presence of a Killing vector.

\section{Acknowledgments}
 We are grateful to A.\ Borowiec of the University of Wroclaw for having
directed our attention to the Chern--Simons formulation of Lovelock
gravity. We  also  thank  L.\ Fatibene  and M.\ Ferraris of the
University of Torino for useful 
discussions and suggestions on the
subject.

%%%%%%%%%%%%%%%%%%%%%%%%%%%%%%%%%%%%%%%%%%%%%%%%%%%%%%%%%%%%%%%%%%%%%%%%%%%%%%%%%%%%%%%%%%%%%%%%


\begin{thebibliography}{99}


\bibitem{Ach} A.Achucarro, P.K.Townsend, Phys.Lett. B
{\bf 180}  (1986), 89.


\bibitem{HIO} G.\ Allemandi, M.\ Francaviglia, M.\ Raiteri:  Class. Quant. Grav. 
{\bf 19} (2002), 2633-2655 (gr-qc/0110104).


\bibitem{CS3d} G.\ Allemandi, M.\ Francaviglia, M.\ Raiteri:  Class. Quant. 
Grav. {\bf 20} (2003), 483-506 (gr-qc/0211098).

\bibitem{Anco} S.C.Anco, R.S.Tung, J. Math. Phys {\bf 43}
(2002), 3984; 

S.C.Anco, R.S.Tung, J. Math. Phys {\bf 43}
(2002), 5531.

\bibitem{btz}  M. Ba\~{n}ados, C. Teitelboim, J. Zanelli: Phys. Rev. Lett.
{\bf  69} (1992), 1849-1851 (hep-th/9204099);\\
M. Ba\~{n}ados, M. Henneaux, C. Teitelboim, J. Zanelli:  Phys. Rev. D
{\bf  48} (1993), 1506-1525 (gr-qc/9302012).

\bibitem{Bana5d}  M. Ba\~{n}ados, C. Teitelboim, J. Zanelli: Phys. Rev. D
{\bf  49} (2) (1994), 975-986 (gr--qc/9307033).

\bibitem{Bana5d1}	M. Ba\~{n}ados: Phys. Rev. D {\bf 65} (2002), 
  (hep-th/0109031); \\
M. Ba\~{n}ados: Nucl. Phys. Proc. Suppl. {\bf 88} (2000), 17-26
  (hep-th/9911150).

\bibitem{Barnich}  G. Barnich, F. Brandt,
 Nucl.Phys. {\bf B633} (2002) 3 (hep--th/0111246).

\bibitem{ct} N.D. Birrel, P. C. W. Davies: {\em Quantum fields in curved space}, Cambridge University Press (1982).
 
\bibitem{Booth}{I. Booth: PhD Thesis from University of Waterloo (gr-qc/0008030);\\ 
I. Booth, R.B. Mann: Phys. Rev. D {\bf
59}, 064021 (gr--qc/9810009); \\I. Booth, R.B. Mann: Phys. Rev. D 124009 ($1999$) 
 (gr--qc/9907072).}




\bibitem{boul} D.G. Boulware, S. Deser: Phys. Rev. Lett. {\bf 55} (24) (1985), 2656-2660.



\bibitem{BY}{J.\ D.\ Brown, J.\ W.\ York, Phys.\ Rev.\ D
{\bf 47} (4), (1993),1407;


J.\ D.\ Brown, J.\ W.\ York, Phys.\ Rev.\ D
{\bf 47} (4) (1993), 1420. }


\bibitem{CF} 
  A. H. Chamseddine:  Nucl.
   Phys. B {\bf 346} (1990), 213-234;\\
  A. H. Chamseddine, J. Fr\"ohlich: Comm. Math. Phys. {\bf 147} (1992), 549.

\bibitem{Nester} C. M. Chen, J. M. Nester: 
Gravitation \& Cosmology {\bf 6}, 257, (2000)
 (gr--qc/0001088).



\bibitem{BHS} J. Crisostomo, R. Troncoso, J. Zanelli: Phys.Rev. D {\bf 62} (2000), 084013 (hep-th/0003271).

\bibitem{deha} M. H. Dehghani: Phys. Rev. D {\bf 67} (2003), 064017  (hep-th/0211191).



\bibitem{dese} S. Deser, B. Tekin: Phys. Rev. D {\bf 67} (2003), 084009
  (hep-th/0212292).


\bibitem{libroFF}{L. Fatibene and M. Francaviglia}, 
{\it  Natural and gauge
natural  formalism for classical field theories: a geometric perspective
including  spinors and gauge fields,}  in Pubblication for Kluwer Academic
Publishers.

\bibitem{Koslor}
L.\ Fatibene, M.\ Ferraris, M.\ Francaviglia, M.\ Godina:
 Gen. Rel. Grav. {\bf 30} (1998),  1371-1389.

\bibitem{Remarks}
L.\ Fatibene, M.\ Ferraris, M.\ Francaviglia, M.\ Raiteri:
 Annals of Phys.
{\bf 275} (1999),  27.

\bibitem{BTZ}{
L.\ Fatibene, M.\ Ferraris, M.\ Francaviglia, M.\ Raiteri, 
Phys. Rev. D{\bf 60}, 124012 (1999);

L.\ Fatibene, M.\ Ferraris, M.\ Francaviglia, M.\ Raiteri,  Phys.
Rev. D{\bf 60}, 124013 (1999);

L.  Fatibene, M.  Ferraris, M.  Francaviglia, M.
Raiteri, Annals of Phys. {\bf 284}, 197 (2000),  gr-qc/$9906114$.}



\bibitem{CADM} M.\ Ferraris and M.\ Francaviglia: Atti Sem. Mat. Univ.
Modena,
{\bf 37} (1989), 61;
M.\ Ferraris and M.\ Francaviglia: Gen.\ Rel.\ Grav., {\bf 22}, (9) (1990),
965;
M. Ferraris, M. Francaviglia, I. Sinicco:  {\em   Il Nuovo Cimento},
$\mathbf{107 B} $, N. $11$, ($1992$),    $1303$; M.  Ferraris, M. 
Francaviglia: \emph{$7$th Italian Conference on  General Relativity  and
Gravitational Physics}, Rapallo (Genoa),  September $3$--$6$, $1986$. 

\bibitem{EM} M.\ Ferraris, M.\ Francaviglia, M.\ Raiteri:  Class. Quant.
Grav. \emph{in press} (gr-qc/0305047).

\bibitem{Lagrange}{M. Ferraris, M. Francaviglia,
in: {\it Mechanics, Analysis and Geometry: 200 Years after Lagrange},
Editor: M. Francaviglia, Elsevier Science Publishers B.V., (Amsterdam,
1991) 451;

A. Trautman, in: {\it Gravitation: An Introduction to
Current Research}, L.
Witten ed.
(Wiley, New York, 1962) 168; 

A. Trautman, Commun. Math. Phys., {\bf  6}, (1967),  248.}

\bibitem{Robutti} M. Ferraris, M. Francaviglia and O. Robutti, 
in:\emph{ Geometrie et Physique},
Proceedings of the \emph{Journees Relativistes 1985} (Marseille, 
$1985$), $112$ -- $125$; Y.
Choquet-Bruhat, B. Coll, R. Kerner, A. Lichnerowicz eds. (Hermann, 
Paris, $1987$). 


\bibitem{forth} M.\ Francaviglia, M.\ Raiteri: Class. Quant. Grav.
{\bf 19}, ($2002$),  237-258 (gr-qc/0107074).


\bibitem{Godina} M.\ Godina, P.\ Matteucci: in press for J. Geom. Phys.
(math.DG/0201235);

 P.\ Matteucci, gr--qc/0201079.

\bibitem{vangoden} M.\ Godina, P.\ Matteucci, J. A. Vickers:  J. Geom. Phys. {\bf 39} (2001),
265-275.

\bibitem{Spencer} H.Goldschmidt, D.Spencer, J. Diff. Geom. 13 (1978) 455.


\bibitem{stringa} M.B. Greens, J.H. Schwarz, E.Witten: {\it Superstring Theory}
(Cambridge University Press, England 1987);\\
D. Lust, S. Theusen: {\it Lectures
on String Theory} (Springer, Berlin, 1989);\\
J.Polchinski: {\it String Theory}
(Cambridge University Press, England 1998).


\bibitem{HEllis}  S. W. Hawking, G.F.R. Ellis, \emph{The Large Scale 
Structure of Space--Time}, (Cambridge University Press, Cambridge 
$1973$).

\bibitem{HawHun}{S.\ W.\ Hawking, C.\ J.\ Hunter, Class. Quant. Grav. {\bf
13}, (1996) 2735; 

S.\ W.\ Hawking, G.\ T.\ Horowitz, Class. Quant. Grav. {\bf
13}, (1996) 1487.}


\bibitem{Wald} V. Iyer and R. Wald: Phys. Rev. D {\bf 50} (1994), 846
(gr-qc/9403028).



\bibitem{Wald95} V. Iyer, R. M. Wald: Phys.Rev. D {\bf 52} (1995),
 4430-4439 (gr-qc/9503052).

\bibitem{T.Jacobson} T.Jacobson, R. Myers: Phys. Rev. Lett. {\bf 70}, 3684-3687 ($1993$) (hep-th/9305016).


\bibitem{Silva} B. Julia, S. Silva: Class. Quant. Grav.
 {\bf 15}, ($1998$), $2173-2215$ (gr-qc/9804029);\\
 B. Julia, S. Silva: 
Class. Quant. Grav. {\bf 17} ($2000$), $4733-4744$ (gr-qc/0005127).


\bibitem{kij} J. Kijowski: Gen. Rel. Grav. {\bf 29} (1997), 307.


\bibitem{kolar} I. Kol\`a\v r, P.W. Michor, J. Slov{\`a}k:  {\em Natural
Operations in Differential Geometry},  Springer--Verlag
  (New York, $1993$).

\bibitem{lanczos} C. Lanczos: Ann. Math. {\bf 39} (1938), 842.

\bibitem{lov} D. Lovelock: J. Math. Phys. {\bf 12} ($1971$), 498;\\
D. Lovelock: J. Math. Phys. {\bf 13} ($1972$),  874.

\bibitem{Gravitation} {C.W. Misner, K.S. Thorne, J.A. Wheeler,
{\it Gravitation} (Freeman, San Francisco,  1973).}

\bibitem{myers} R. C. Myers: Phys. Rev. D {\bf 36} (1987), 392.


\bibitem{Racz} I. Racz, R. M. Wald: (gr-qc/9507055).

\bibitem{RT} T.\ Regge, C.\ Teitelboim:  Annals of Physics {\bf 88} (1974),
286.

\bibitem{Sarda} G. Sardanashvily: {\em Generalized Hamiltonian 
Formalism for Field Theory}, World
Scientific (Singapore, $1995$).
  
\bibitem{SilvaCS} S. Silva,
Nucl. Phys. {\bf B 558},
 ($1999$), $391$
(hep--th/9809109).



\bibitem{TZ} C.\ Teitelboim, J. Zanelli: in 
\emph{Constraints Theory and Relativistic Dynamics},
Proceedings of the Workshop held in Florence, Arcetri, Italy (World
Scientific, Singapore  1987), 94.

\bibitem{Towsend} P. K. Townsend: {\em Black Holes}, DAMPT University of Cambridge.

\bibitem{visser} M. Visser: Phys. Rev. D {\bf 48} (1993), 5697-5705 (hep-th/9307194).

\bibitem{Waldbook} R. M. Wald: {\it General Relativity}, University of Chicago Press (Chicago,
$1984$).


\bibitem{Wi1}
E. Witten:
 Nucl. Phys. B {\bf 311} (1988), 96;\\
E. Witten:
Nucl. Phys. B {\bf 323} (1989), 113.


\end{thebibliography}
\end{document}